\documentclass[aps,superscriptaddress,twocolumn,pre,longbibliography,floatfix]{revtex4-1}

\usepackage{amsmath,amsthm,amsfonts,amssymb,bm,graphicx,color,mathpazo,times, braket}
\usepackage[colorlinks={true}, citecolor={blue}, filecolor={blue}, linkcolor={blue}, urlcolor={blue}]{hyperref}
\usepackage[caption=false]{subfig}
\usepackage{graphicx}
\usepackage{graphicx}
\usepackage{soul}
\usepackage{amsmath}
\usepackage{soul}
\usepackage{academicons}

\allowdisplaybreaks

\begin{document}
\title{Quantum thermometry for ultralow temperatures using probe and ancilla qubit chains}
\author{Asghar Ullah}
 \email{aullah21@ku.edu.tr}
\affiliation{Department of Physics, Ko\c{c} University, 34450 Sar\i yer, Istanbul, T\"urkiye}
\author{Vipul Upadhyay }
\email{vipuupadhyay4@gmail.com}
\affiliation{Department of Chemistry, Institute of Nanotechnology and Advanced Materials, Center for Quantum Entanglement Science and Technology, Bar-Ilan University, Ramat-Gan, 52900, Israel}

\author{\"Ozg\"ur E. M\"ustecapl\i o\u glu}	
	\email{omustecap@ku.edu.tr}
	\affiliation{Department of Physics, Ko\c{c} University, 34450 Sar\i yer, Istanbul, T\"urkiye}
	\affiliation{T\"UB˙ITAK Research Institute for Fundamental Sciences, 41470 Gebze, T\"urkiye}
 \date{\today} 
\begin{abstract}
We propose a scheme to enhance the range and precision of ultralow temperature measurements by employing a probe qubit coupled to a chain of ancilla qubits. Specifically, we analyze a qubit chain governed by Heisenberg $XX$ and Dzyaloshinskii-Moriya (DM) interactions. The precision limits of temperature measurements are characterized through the evaluation of quantum Fisher information (QFI). Our findings demonstrate that the achievable precision bounds, as well as the number of peaks in the QFI as a function of temperature, can be controlled by adjusting the number of ancilla qubits and the system's model parameters. These results are interpreted in terms of the influence of energy transitions on the range and the number of QFI peaks as a function of temperature. This study highlights the potential of the probe qubit-ancilla chain system as a powerful and precise tool for quantum thermometry in the ultralow temperature regime.
\end{abstract}	
\maketitle
\section{Introduction}
Precise measurements of ultracold temperatures are essential for effective heat management, ensuring the robust and optimal operation of quantum devices ~\cite{Goold_2016,DePasquale2018,paris2009,PhysRevX.10.011018,Kucsko2013}. The rapidly developing field of quantum thermometry addresses fundamental and practical questions related to temperature measurements and precision limits across a wide range of quantum technology platforms, including solid-state impurities~\cite{PhysRevA.107.042614, Mihailescu_2024, PhysRevA.110.032611,PhysRevResearch.4.023191}, optomechanical systems~\cite{PhysRevA.92.031802}, quantum gases~\cite{PhysRevLett.122.030403, PhysRevLett.128.040502}, dephasing impurities~\cite{PhysRevLett.125.080402,PhysRevA.109.023309}, and topological spinless fermions~\cite{srivastava2023topologicalquantumthermometry}.
Optimal quantum thermometry schemes have been investigated in various contexts, such as probe optimization~\cite{Mukherjee2019,PRXQuantum.3.040330}, critical quantum thermometry~\cite{Aybar2022criticalquantum,srivastava2023topologicalquantumthermometry,PhysRevA.97.063619,Mihailescu_2024}, sequential measurement thermometry~\cite{PhysRevLett.129.120503, Burgarth_2015, PhysRevApplied.22.024069}, and global quantum thermometry frameworks~\cite{PhysRevLett.127.190402,Mok2021,PhysRevResearch.6.L032048,PhysRevResearch.6.043171}. Beyond conventional equilibrium measurements, non-equilibrium approaches to extract temperature information have also been explored~\cite{PhysRevA.99.062114, PhysRevA.108.032220, PhysRevApplied.17.034073, PhysRevA.109.L060201,PhysRevE.110.024132}. These include quantum thermometry techniques based on repeated interactions between a probe and the system~\cite{PhysRevLett.123.180602, PhysRevA.102.042417, PhysRevA.105.012212,PhysRevLett.123.180602}, as well as periodically driven or modulated probes~\cite{PhysRevX.10.041054, PhysRevA.108.022608}. While typical quantum thermometers are based on two-level quantum systems (qubits)~\cite{PhysRevA.110.032611,Burgarth_2015,PhysRevA.91.012331,PhysRevA.84.032105,Razavian2019,PhysRevA.86.012125,Connor_2024} in diverse configurations \cite{PhysRevX.10.011018,Kucsko2013,PhysRevA.93.043607,PhysRevLett.129.120404,Halbertal2016,Karimi2020}, recent studies have highlighted the advantages of multi-level systems, such as a two-level system with a degenerate excited state, for enhancing both range and precision \cite{Mehboudi_2015,Campbell_2018}. The question of finding a more experimentally feasible system with high sensitivity to a broader range of ultralow temperatures is critical for the efficiency of quantum technologies. 

Similar enhancements in quantum thermometry achieved with highly degenerate two-level systems can also be realized using multiple qubit systems. An exceptional opportunity in quantum thermometry with multiple qubits lies in leveraging quantum correlations~\cite{porto2024enhancinggaussianquantummetrology,PhysRevLett.128.040502,Ullah_2025}, entanglement~\cite{PhysRevLett.114.220405, TIAN20171, PhysRevResearch.2.033498}, squeezing~\cite{Mirkhalaf_2024}, and coherence~\cite{PhysRevA.110.032605, Ullah2023, doi:10.1073/pnas.1306825110} to enhance thermometric precision~\cite{PhysRevLett.125.080402,PhysRevLett.118.130502, PhysRevResearch.2.033497, PhysRevA.96.012316, PhysRevA.98.050101, PhysRevA.98.050101, Potts2019fundamentallimits, Campbell_2017}. Ancilla-assisted quantum probes can be used to enhance the sensitivity of single temperature estimation~\cite{Zhang2022,PhysRevA.98.042124,PhysRevA.109.042417,PhysRevA.110.062406}. Using a single probe qubit coupled to a network of ancilla qubits allows for estimating multiple temperatures~\cite{Ullah2023,PhysRevA.110.032605}. While promising, these approaches face notable limitations, such as the need for quantum measurements due to the presence of coherences in the probe qubit state. Moreover, the scheme in Ref.~\cite{Ullah2023} relies on an asymmetric interaction, where the $z$-component of one spin is coupled to the $x$-component of another spin. This type of interaction has not yet been experimentally realized and poses significant challenges due to its inherent asymmetry. Interestingly, there exists a naturally occurring weak asymmetric spin-spin interaction, the DM interaction~\cite{DZYALOSHINSKY1958241, Moriya}, which emerges alongside the exchange interaction in systems lacking inversion symmetry~\cite{PhysRevResearch.2.033092, PhysRevA.84.042302}. In this work, we consider a probe qubit coupled to a chain of $N-1$ ancilla qubits, where all the qubits interact through both the DM interaction~\cite{DZYALOSHINSKY1958241, Moriya} and the Heisenberg $XX$ exchange interaction.

To characterize the precision limits of spin-chain ancilla-assisted ultracold temperature estimation, we evaluate the QFI~\cite{paris2009}. We begin by examining the simplest case of two coupled qubits and demonstrate that the presence of two different energy channels (allowed transition frequencies) enables sensitivity of the scheme to two different temperatures, leading to a second peak in the QFI. Moreover, the precision associated with these QFI peaks can be finely tuned by adjusting the relevant system parameters. Building on this foundation, we extend our analysis to systems with multiple ancilla qubits. Remarkably, the system parameters can be determined to ensure that each additional ancilla qubit to the spin chain introduces a distinct energy channel, further enhancing the range of temperature estimations and, most remarkably, generating additional peaks in the QFI at progressively lower temperatures. For a spin chain of $N$ qubits, the number of possible energy transitions increases exponentially with $N$. To maximize the utility of these transitions, we carefully adjust the system parameters to make the transition energies as distinct as possible, enabling the probe qubit to provide sensitive temperature information across a broader range of ultralow temperatures. In a system with Heisenberg XX and DM interactions, we demonstrate that the maximum number of distinct transitions is limited to $N$ for a chain of $N$ qubits. Consequently, our scheme achieves a linear increase in the number of measurable transitions as the system size grows, offering an efficient pathway to expand the operational range and precision of quantum thermometry.

Compared to previous quantum thermometry proposals, we summarize the distinct features of our scheme: (i) Our probe and ancilla qubit chain exhibits high sensitivity for more than one temperature, as indicated by multiple peaks in the QFI that scale with the number of ancilla qubits, which typically requires highly degenerate multilevel probes~\cite{Campbell_2018}. (ii) Our scheme is based on complete thermalization of the spin chain, making it autonomous in comparison to other proposals for wide-range probe thermometers~\cite{Glatthard2022bendingrulesoflow, Mukherjee2019}. (iii) It does not require initial coherence~\cite{PhysRevA.98.042124} or an initial entangled state~\cite{PhysRevResearch.2.033497,Tham2016} to enhance precision. (iv) Our results are robust against the bath spectra, which was not the case for previous multiple-peaked QFI probes~\cite{Mukherjee2019}. (v) While critical quantum thermometry offers enhanced precision due to quantum phase transitions~\cite{Jepsen2020,Salado-Mejía_2021,PhysRevResearch.6.043094}, it faces limitations in operating effectively over a wide temperature range, especially at low temperatures.
The remainder of the paper is organized as follows. In Sec.~\ref{mod}, we introduce our theoretical model for temperature estimation and discuss the open system dynamics, with the Gibbs thermal state serving as the steady state of the entire system. In Sec.~\ref{res}, we present our findings for the case of a two-qubit system, focusing on populations, energy transitions, and the QFI analysis, with additional details on the two-qubit state provided in Appendix~\ref{trans}. Sec.~\ref{mq} extends these results to systems with multiple ancilla qubits. Additionally, classical Fisher information and optimal measurement strategies are discussed in Appendix~\ref{QFI}, while Appendix~\ref{Nqubits} presents the theoretical determination of the spin chain density matrix. Finally, the paper is concluded in Sec.~\ref{conc}.
\section{Model}\label{mod}
\begin{figure}[t!]
    \centering
    \includegraphics[scale=0.75]{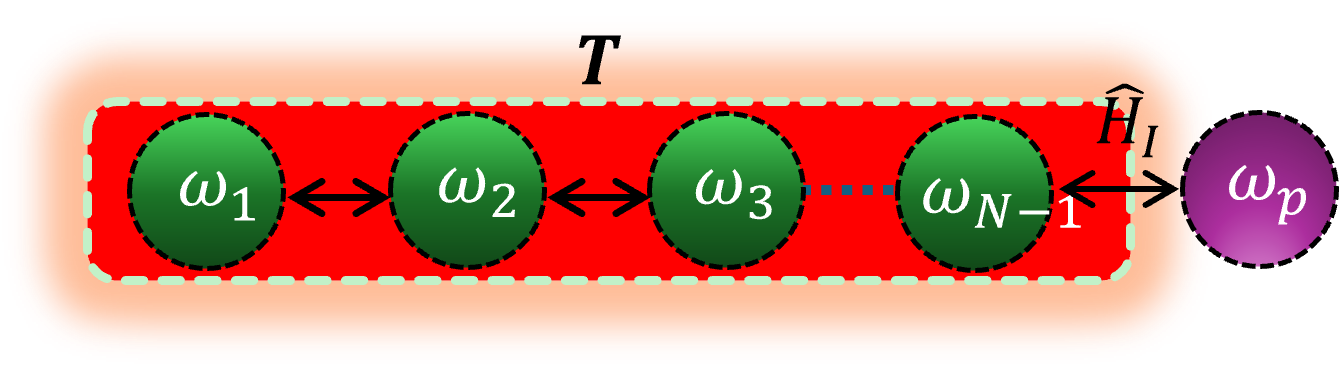}
    \caption{A schematic representation of our thermometry scheme is shown. The system consists of a probe qubit with a transition frequency $\omega_p$ located outside of the thermal sample. This qubit is used as a probe to measure the unknown temperature $T$ of a sample. The measurement is facilitated by ancilla qubits with transition frequencies $\omega_i$ ($i=1,2,3,\ldots,N-1$), which are immersed in the sample. The qubits are coupled via a combination of the Heisenberg $XX$ interaction ($\hat{H}_{XX}$) with coupling strength $J$, and DM interaction ($\hat{H}_{DM}$) characterized by strength $g$. }
    \label{fig1}
\end{figure}

Our quantum thermometry model is based on a physical system of $N$ spin-$1/2$ qubits coupled through Heisenberg $XX$ and DM interactions. To measure the temperature of a thermal sample, $N-1$ ancilla qubits are immersed in a thermal bath, with one qubit remaining isolated to serve as a probe for estimating the unknown temperature $T$, as illustrated in Fig.~\ref{fig1}. We begin with a brief discussion of the two-qubit system and then extend our analysis to the case involving multiple ancilla qubits. The total Hamiltonian for the two-qubit system is given by the following~\cite{Micadei2019,PhysRevA.84.042302,PhysRevResearch.2.033092}
\begin{equation}\label{Model}
\begin{aligned}
\hat{H}_S=&\hat{H}_0+\hat{H}_{I}\\
    =&\hat{H}_0+\hat{H}_{XX}+\hat{H}_{DM},
    \end{aligned}
\end{equation}
where $\hat{H}_0$ is the Hamiltonian of noninteracting qubits and is given by
\begin{equation}\label{eq1}
    \hat{H}_{0}=\frac{1}{2}\omega_{a}\hat{\sigma}^z_{a}\otimes\hat{I}_p+\frac{1}{2}\omega_{p}\hat{I}_a\otimes\hat{\sigma}^z_{p},
\end{equation}
where $\omega_a$ and $\omega_p$, represent the frequencies of the ancilla and probe qubit, respectively, and $\hat{\sigma}^\alpha_i$ represent the $\alpha=x,y,z$ components of the Pauli spin-1/2 operators for the probe ($i=p$) and ancilla qubit ($i=a$). The unitary operators are denoted by $\hat{I}_i$.
The second term in Eq.~(\ref{Model}) is the Heisenberg $XX$ coupling between the two qubits and is given by
\begin{equation}
    \hat{H}_{XX}=J(\hat{\sigma}^x_a\hat{\sigma}^x_p+\hat{\sigma}^y_a\hat{\sigma}^y_p),
\end{equation}
where $J$ denotes the coupling strength of the exchange interaction.
The last term in Eq.~(\ref{Model}) is the DM interaction, which is an antisymmetric exchange interaction described by the Hamiltonian $\hat{H}_{DM}$, which reads as
\begin{equation}\label{int}
    \hat{H}_{DM}=g(\hat{\sigma}^x_a\hat{\sigma}^y_p-\hat{\sigma}^y_a\hat{\sigma}^x_p).
\end{equation}
The strength of the DM anisotropic field, which is taken along the $z$-direction, is characterized by $g$.

The free Hamiltonian of the thermal sample consists of an infinite number of non-interacting bosonic modes and is given by
\begin{equation}
    \hat{H}_B=\sum_k\omega_{k}\hat{b}^\dagger_{k}\hat{b}_{k},
\end{equation}
where $\hat{b}^\dagger_k(\hat{b})$ are the creation (annihilation) operators of $k$-th mode in the bath with frequency $\omega_k$. The system-bath interaction is described by the Hamiltonian
\begin{equation}\label{sb}
    \hat{H}_{SB}=\hat{\sigma}^x_a\otimes\sum_{k}s_{k}(\hat{b}_k+\hat{b}^\dagger_{k}),
\end{equation}
where $\hat{\sigma}_a^x$ is the Pauli spin $X$ operator for the ancilla qubit and $s_{k}$ are the coupling coefficients between the ancilla qubit and each $k$th mode of the sample.

\begin{figure}[t!]
    \centering
    \includegraphics[scale=0.6]{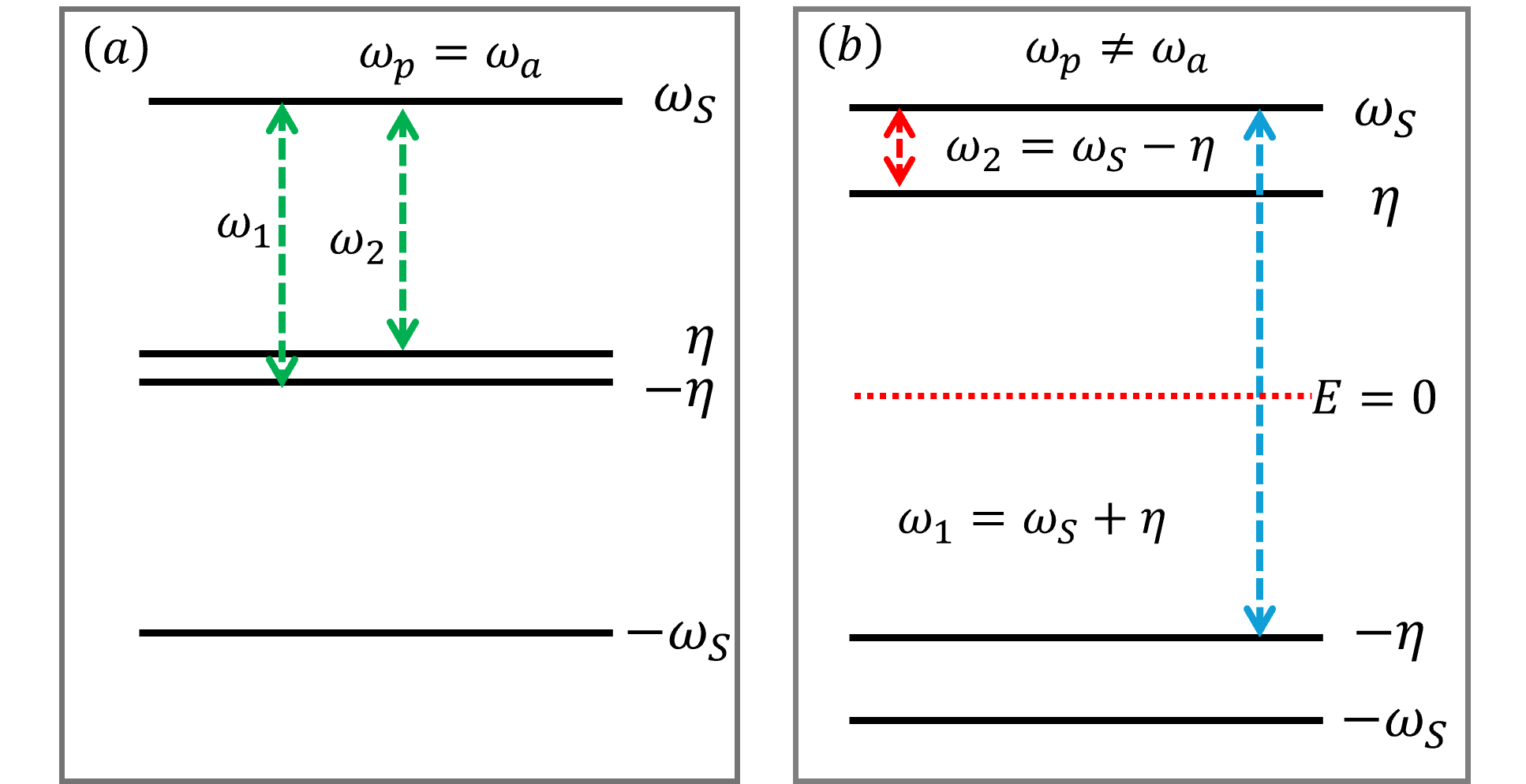}\\
    \caption{(a) Energy transitions induced by the bath when the two qubits are resonant, such as $\omega_p=\omega_a$. The two transitions $\omega_1$ and $\omega_2$ are of almost the same order, such as $\omega_1\sim\omega_2$. (b) shows the transitions induced when the two qubits are off-resonant, such as ($\omega_p\neq\omega_a$). In this case, the transition energies $\omega_1$ and $\omega_2$ are of different orders.}
    \label{fig2}
\end{figure}

The eigenvalues of the system Hamiltonian~(\ref{Model}) are $\pm\omega_S$ and $\pm\eta$, where
\begin{equation}
    \begin{aligned}
        \omega_S=&\frac{\omega_p+\omega_a}{2},\quad \text{and} \quad \eta =\sqrt{\omega^2_D+4g^2+4J^2} 
    \end{aligned}
\end{equation}
with $\omega_D:=(\omega_p-\omega_a)/2$ as a notation for the
sake of convenience. The eigenvectors associated with the eigenvalues can be expressed as
\begin{equation}
    \begin{aligned}
        |1\rangle:&=|\omega_S\rangle=|00\rangle,\\
        |2\rangle:&=|\eta\rangle=\frac{2(J+ig)}{\Delta}|01\rangle+\frac{\eta-\omega_D}{\Delta}|10\rangle,\\
        |3\rangle:&=|-\eta\rangle=-\frac{\eta-\omega_D}{\Delta}|01\rangle+\frac{2(J-ig)}{\Delta}|10\rangle,\\
        |4\rangle:&=|-\omega_S\rangle=|11\rangle,
    \end{aligned}
\end{equation}
where the parameter $\Delta$ is defined as
\begin{equation}
    \Delta:=\sqrt{4J^2+4g^2+(\omega_D-\eta)^2}.
\end{equation}
 The energy transitions induced by the bath are depicted in Fig.~\ref{fig2}, which displays all the possible transitions for $\omega_a=\omega_p$ (Fig.~\ref{fig2}(a)) and $\omega_a\neq\omega_p$ (Fig.~\ref{fig2}(b)). The system Hamiltonian~(\ref{Model}) in the diagonal basis takes the form of $\tilde{H}_s=\text{diagonal}(\omega_S,\eta,-\eta,-\omega_S)$.
The Hamiltonian in Eq.~\eqref{Model} can be realized across various experimental platforms. For example, trapped ion registers~\cite{Blatt2012, Richerme2014, PhysRevLett.92.207901, PhysRevLett.111.040601} and superconducting qubits~\cite{Wang2019, Barends2016} already support qubit chains, making them promising candidates for temperature monitoring. Ultracold atoms~\cite{Jepsen2020,Galitski2013,Bloch2012} in optical lattices simulate spin models via synthetic gauge fields, while solid-state systems~\cite{Fert2013,PhysRevA.57.120}, including quantum dots and NV centers, naturally exhibit Heisenberg along with the spin-orbit couplings. Nuclear magnetic resonance experiments~\cite{Micadei2019,PhysRevA.79.012305} provide a robust testbed for DM-coupled qubits, and quantum dots coupled through exchange and DM interactions~\cite{PhysRevB.64.075305,PhysRevB.77.193306,PhysRevLett.110.046803} may have potential applications in temperature sensing.
\section{Results}\label{res}
\subsection{Sensitivity of populations with $T$}
The density matrix of the probe qubit is found by tracing out the ancillary qubit from a two-qubit density matrix (see Appendix~\ref{trans} for more details). Consequently, we have
\begin{equation}\label{probe}
    \hat{\rho}_p=\text{Tr}_a[\hat{\rho}_{ss}]=\left(
\begin{array}{cc}
 p(T) & 0 \\
 0 & 1-p(T) \\
\end{array}
\right),
\end{equation}
where the population of the excited state of the probe density matrix is given by
\begin{equation}\label{pop-exact}
\begin{aligned}
p(T)=  \frac{\cos (2 \theta ) \sinh \left(\frac{\eta }{T}\right)+\chi-\sinh \left(\frac{\omega_S}{T}\right)}{2 \chi},
\end{aligned}
\end{equation}
where the parameter $\chi$ is explicitly given in Appendix~\ref{trans}. In any quantum metrological analysis, identifying the parameter points where the density matrix shows the most rapid variation is crucial. Since the probe density matrix in Eq.~\eqref{probe} depends solely on the single independent parameter $p(T)$, our discussion focuses on determining the temperature regions where this function changes most rapidly.
\begin{figure}[t!]
    \centering
    \includegraphics[scale=0.65]{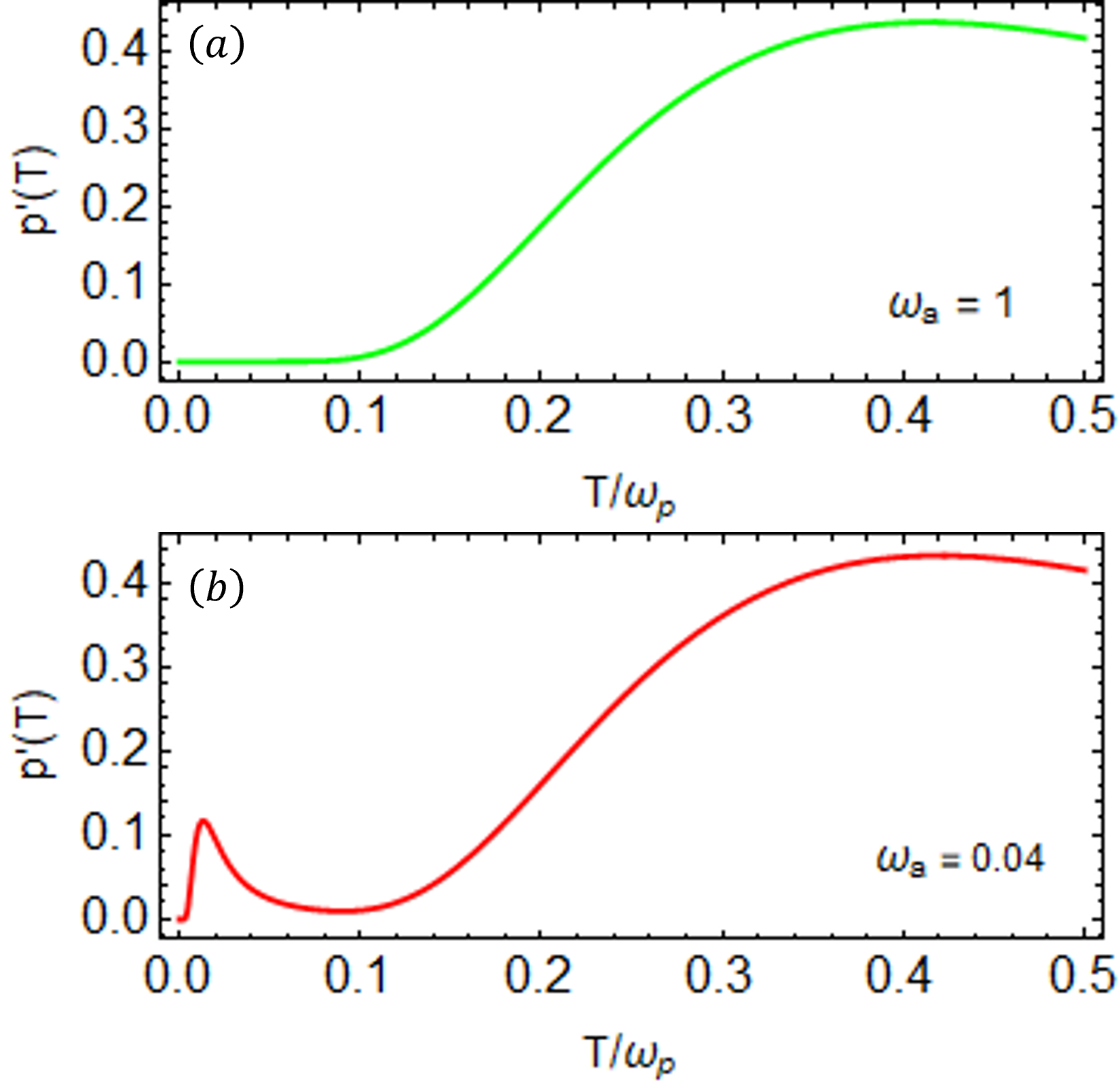}
    \caption{The behavior of the first derivative of population  $p^\prime(T)$ of the excited state of the probe qubit  as a function of temperature $T$ for (a) $\omega_a=1$ and (b) $\omega_a=0.04$. We can see that there is an additional peak at lower temperatures when $\omega_p\neq\omega_a$, while only one peak can be seen for the resonant qubits, $\omega_p=\omega_a$. 
    The parameters are set to $\omega_p=1$, $g=0.02$, and $J=0.04$. All the system parameters are scaled with the probe qubit frequency $\omega_p=1$.}
    \label{fig3}
\end{figure}
All parameters in this work are scaled by the frequency $\omega_p$ and are presented in dimensionless form unless explicitly specified otherwise. We show the exact first derivative of population $p(T)$ (Eq.~(\ref{pop-exact})) of the excited state of the probe as a function of temperature $T$ for two different cases in Fig.~\ref{fig3}. We observe a single temperature peak in the $p^\prime(T)$ of the probe state shown in Fig.~\ref{fig3}(a) when the system exhibits similar frequencies ($\omega_a=\omega_p$). This can be explained by the allowed transitions shown in Fig.~\ref{fig2}(a) offering a single channel to imprint the temperature information. On the other hand, we observe that when the two qubits have different frequencies, it leads to special temperature peaks in $p^\prime(T)$, as shown in Fig.~\ref{fig3}(b). These two peaks in the first derivative of populations can be understood by the allowed transitions shown in Fig.~\ref{fig2}(b), where we can see two distinct energy channels. Based on this analysis, we proceed to a detailed investigation of the probe state’s density matrix using low- and high-temperature approximations to better understand the observed peaks in the population dynamics.
\subsubsection{Low-temperature peak}
The steady-state solution of the system is sensitive to changes in the unknown value of $T$ and this sensitivity can be used to determine the value of $T$. As the system does not generate coherences in the probe state, the populations within the density matrix serve as effective indicators for measuring $T$. Therefore, we will carry out a detailed analysis of the population dynamics of the probe qubit.

\begin{figure}[t!]
    \centering
    \includegraphics[scale=0.64]{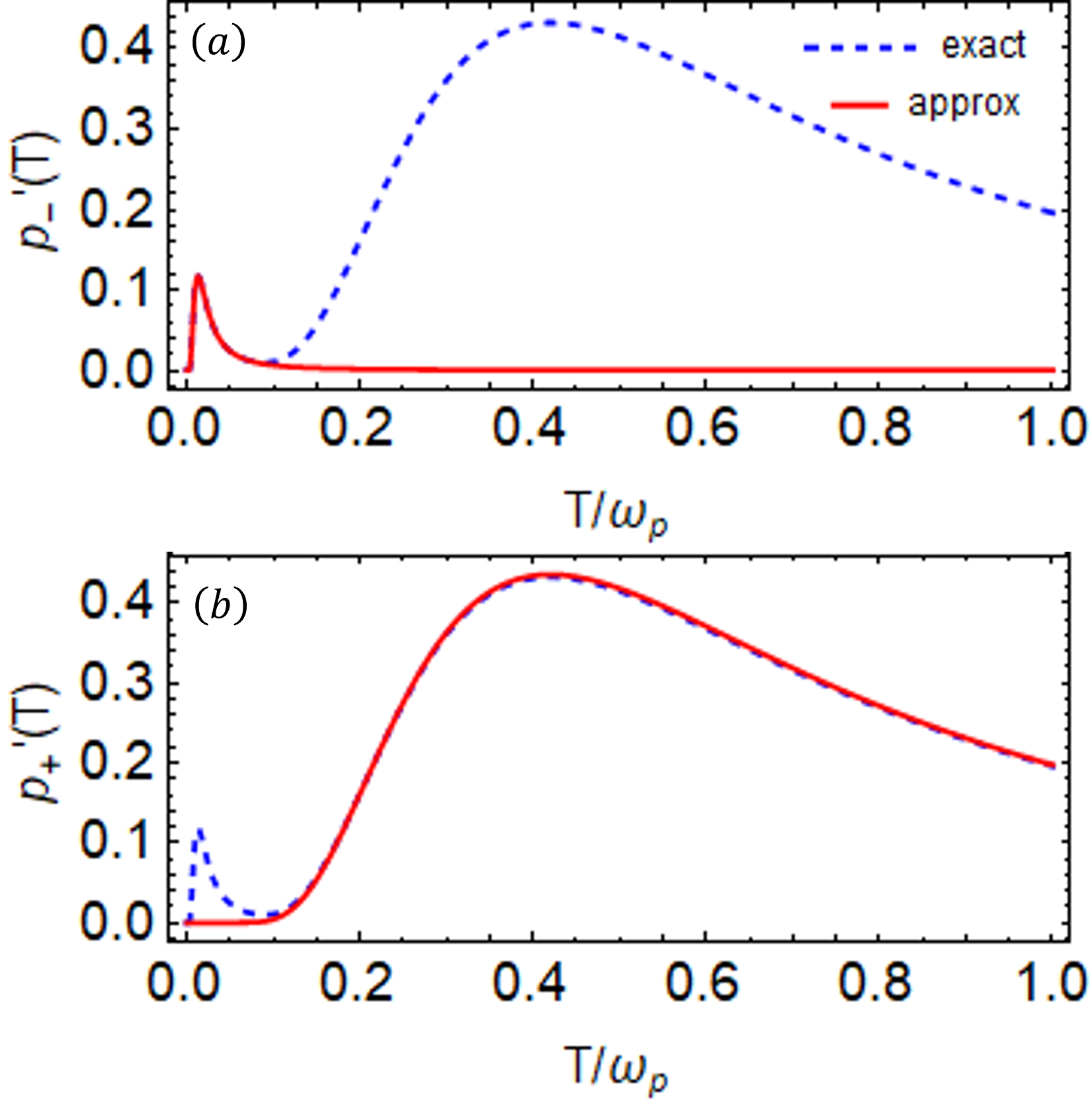}
   \caption{(a) The first derivative of the population \(p'_-(T)\) as a function of \(T\) in the low-temperature limit, while (b) shows the first derivative of the population \(p'_+(T)\) as a function of \(T\) in the high-temperature limit. In both plots, the blue dashed curve represents the exact expression for the first derivative of \(p\) as given in Eq.~(\ref{pop-exact}), while the solid red curve corresponds to the approximate expressions given in Eqs.~(\ref{pLd}) and~(\ref{pHd}), respectively. The parameters are set to \(\omega_p=1\), \(\omega_a=0.04\), \(g=0.02\), and $J=0.04$. All the system parameters are scaled with the probe qubit frequency $\omega_p=1$.}

    \label{fig4}
\end{figure}
We first examine the case where the two qubits are off-resonant and the two transition frequencies are of different orders, specifically, $\omega_p\gg\omega_a$. This means that we want to investigate a very low-temperature regime, and we assume $\{\eta,\omega_S\}\gg T$. Under this assumption, we can write
\begin{equation}
    \cosh(\frac{\eta}{T})\approx\sinh(\frac{\eta}{T})\approx \frac{e^{\eta/T}}{2}.
\end{equation}
Therefore, the probe population simplifies to
\begin{equation}\label{popL}
    p_-(T)\approx \frac{\cos^2\theta}{1+e^{\omega_-/T}},
\end{equation}
where we define $\omega_-=\omega_S-\eta$. The first derivative of population under this limit is given by
\begin{equation}\label{pLd}
    p_-^\prime(T)=\frac{\omega_-\cos ^2(\theta ) \text{sech}^2(\frac{\omega_-}{2 T})}{4 T^2}.
\end{equation}
In Fig.~\ref{fig4}(a), the exact (blue dashed) and approximate (red solid) first derivatives of the population $p(T)$ of the probe are plotted as a function of bath temperature $T$. It shows that the approximation we consider is valid for low-temperature regimes. The derivative of population in Fig.~\ref{fig3}(b) shows that the population of the probe has two peaks, one at low temperature and the other at higher temperature. The additional peak at low temperatures appears due to the fact that the two qubits have a large frequency difference ($\omega_p\gg\omega
_a$), allowing for another energy transition. This figure clearly shows that the two peaks at low temperature from the exact and approximate values of $p$ are consistent.\\

The two peaks in the behavior of the population can be well explained by the transition frequencies of qubits. The possible energy transitions for the qubits are shown in Fig.~\ref{fig2}. When the qubits are resonant, the population value is low in the low-temperature regime, while for the off-resonant qubits, the populations increase rapidly as more transitions occur, indicating two peaks in the population. To investigate the behavior of the lower peak, we solve for the value of the temperature, where the value of $dp/dT$ is maximum such that $d^2p/dT^2=0$.
After some straightforward calculations, the temperature $T^{*}$ at which the change in population `$p$' is maximum is given by a transcendental equation
\begin{equation}\label{T-low}
    T^{*}_\mathrm{low}=\frac{\omega_-}{2}\tanh(\frac{\omega_-}{2T^*}).
\end{equation}

The above expression (\ref{T-low}) clearly shows that the position of the lower peak can be adjusted by modifying the value of \( \omega_- \). This implies the existence of a distinct frequency, $\omega_-$, associated with the high-temperature peak. To elucidate the physical mechanism underlying the two peaks in the population, we consider two temperature exchange channels corresponding to the transitions at $\omega_{+}$ and $\omega_-$. Consequently, for two peaks to manifest in $p'(T)$, there must be a significant gap between the magnitudes of $\omega_+$ and $\omega_-$.
\subsubsection{High-temperature peak}
To resolve the peak at high temperature, we apply certain assumptions, specifically setting $\cos2\theta\approx-1$ for the parameters under consideration. Under such assumptions, the expression of $p$ in Eq.~(\ref{pop-exact}) reduces to
\begin{equation}\label{popH}
    p_+(T)\approx\frac{1}{1+e^{\frac{\omega_+}{T}}},
\end{equation}
where $\omega_+=\omega_S+\eta$. The first derivative of $p$ is given as
\begin{equation}\label{pHd}
    p_+^\prime(T)=\frac{\omega_+ \text{sech}^2(\frac{\omega_+}{2 T})}{4 T^2}.
\end{equation}
This reveals another effective frequency $\omega_+$ associated with the peak at higher temperatures. Since we are examining the first derivative of population $p$, therefore we need to maximize the first derivative of $p$
After doing some algebra, we get the value of $T^*$ at which the population has a maximum value, which is given as
\begin{equation}
    T^*_\text{high}=\frac{\omega_+}{2}\tanh(\frac{\omega_+}{2T^*}).
\end{equation}
In Fig.~\ref{fig4}(b), we present the first derivative of the population as a function of temperature \( T \) for off-resonant qubits using the approximate above expression~\eqref{pHd}. The approximate results at high temperatures align closely with the exact behavior of \( dp/dT \). The two energy channels, corresponding to the effective frequencies \( \omega_+ \) and \( \omega_- \), play a crucial role in resolving the two peaks. These peaks emerge when there is a significant difference between \( \omega_+ \) and \( \omega_- \) and it allows more transitions. Further discussions are provided in the next section.

\subsection{QFI analysis}
The sensitivity of the density matrix to small changes in the unknown parameter $T$ can be quantified using the QFI. Specifically, for the density matrix of a two-level system, the QFI is calculated as follows~\cite{PhysRevA.87.022337, Dittmann_1999}
\begin{equation}
    \mathcal{F}_Q(T)=\text{Tr}\big[\left(\partial_T \rho_T\right)^2\big]+\frac{1}{|\rho_T|}\text{Tr}\big[(\rho_T\partial_T \rho_T)^2\big],\label{qfi}
\end{equation}
where $\partial_T:=\partial/\partial_T$ represents the partial derivative with respect to $T$ and $|\rho_T|$ shows the determinant of the density matrix $\rho_T$. In a single-shot measurement scenario (such as $m=1$, where $m$ shows the number of measurements performed), the ultimate precision limit is determined by the Cramér-Rao bound, which relates the variance in $T$ with the inverse of QFI through relation $\text{Var}(T)\ge 1/\mathcal{F}_T$. Using the above formula, we find the exact QFI for the probe qubit, which is given below
    \begin{equation}
        \mathcal{F}_Q(T)=\frac{\left(B+\zeta -\eta  \cos (2 \theta )+\text{$\omega $s}\right)^2}{T^4 \chi ^2 \left[\chi ^2-\left(\sinh \left(\frac{\text{$\omega $s}}{T}\right)-\cos (2 \theta ) \sinh \left(\frac{\eta }{T}\right)\right)^2\right]},\label{exact}
    \end{equation}
where
\begin{equation}
\begin{aligned}
    \mathcal{\mathcal{B}}:=&\sinh \left(\frac{\eta }{T}\right) \sinh \left(\frac{\text{$\omega $s}}{T}\right) (\text{$\omega $s} \cos (2 \theta)-\eta),\\
    \zeta&:=\cosh \left(\frac{\eta }{T}\right) \cosh \left(\frac{\text{$\omega $s}}{T}\right) (\text{$\omega $s}-\eta  \cos (2 \theta )),
    \end{aligned}
\end{equation}
\begin{figure}[t!]
    \centering
    \includegraphics[scale=0.78]{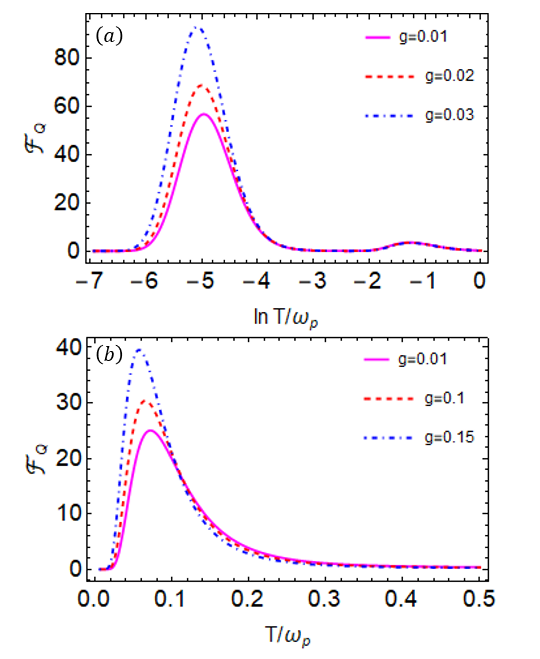}
    \caption{(a) QFI $\mathcal{F}_Q$ as a function of an unknown parameter $T$ for different values of coupling strength $g$ at $\omega_p=1$ and $\omega_a=0.04$. The solid magenta, orange dashed, and blue dot-dashed curves correspond to $g=0.01$, $g=0.02$,  and $g=0.03$, respectively. We set $J=0.05$. (b) QFI for the resonant qubits case, such as $\omega_p=\omega_a=1$. The solid magenta, orange dashed, and blue dot-dashed curves correspond to $g=0.01$, $g=0.1$, and $g=0.15$, respectively. Here, we set $J=0.35$. All the system parameters are scaled with the probe qubit frequency $\omega_p=1$.}
    \label{fig5}
\end{figure}
where $\mathcal{B}$ and $\zeta$ in the above Eq.\eqref{exact} are defined the sake of bravity. The QFI as a function of $T$ for different values of coupling strength $g$ is plotted in Fig.~\ref{fig5}. In particular, Figs.~\ref{fig5}(a) and ~\ref{fig5}(b) depict two distinct cases: one where QFI exhibits two peaks as a function of temperature $T$, while the other shows a single peak in QFI. We can now have one-to-one correspondence between the first derivative of population and QFI for off-resonant qubits. We compare Figs.~\ref{fig3}(b)  and~\ref{fig5}(a), we observe that similar to the first derivative of $p(T)$, the QFI also exhibits a second peak at low temperature. Notably, the height of the additional peak in the QFI at lower temperatures can be enhanced by increasing the coupling strength $g$, which consequently improves the precision of temperature estimation. Nevertheless, the precision at low temperatures can be further enhanced by adjusting the coupling parameter \( J \) while keeping \( g \) fixed.
The position and magnitude of this peak can be understood through the population expression in Eq.~(\ref{popL}). As shown, the numerator contains a $\cos^2\theta$ term that depends on the coupling parameters $g$ and $J$ and this term increases with the value of $g$ or $J$, while the denominator primarily influences the location of the peak. Thus, by tuning $g$ $J$, we can vary the height of the QFI peak without shifting its position significantly. This provides a useful mechanism for optimizing thermometric precision. In contrast, the peak at higher temperatures remains unaffected by the value of these coupling parameters, as the population in Eq.~(\ref{popH}) does not exhibit any dependence on $g$ or $J$. The high-$T$ peak is identical to the QFI of a two-level system at thermal equilibrium~\cite{Ullah2023}. This shows that the low-temperature peak can be adjusted by tuning \( g \), while the high-temperature peak remains stable, as it depends on other system parameters. We can conclude from these results that adding more ancilla qubits may increase the number of peaks in the QFI. Similar observations with multiple peaks in QFI have been reported for periodically driven probes [36], though the physical mechanism behind multiple peaks in the QFI is fundamentally different in these non-autonomous schemes.
\subsubsection{Role of transition energies}
The appearance of two peaks in the QFI can be explained by the presence of two energy channels, $\omega_{-}$ and $\omega_{+}$ appearing in Eqs.~\eqref{pLd} and~\eqref{pHd}, respectively. When the two qubits are resonant, such that $\omega_a = \omega_p = 1$, and with coupling strengths of $g = 0.03$ and $J=0.05$, we find that $\omega_{-} = 0.88$ and $\omega_{+} = 1.11$. This shows that the two frequencies, $\omega_{-}$ and $\omega_{+}$, are relatively close but not identical, with a difference of $0.23$. This means that these two energy channels do not allow efficient temperature information transfer in the lower temperature regime. As a result, we observe only a single peak in the QFI, as shown in Fig.~\ref{fig5}(b). However, the precision of temperature estimation due to this peak can be enhanced by increasing the value of $g$ and $J$.

In contrast, when we consider an off-resonant case, where $\omega_a = 0.04$ and $\omega_p = 1$ with the same coupling strengths $g$ and $J$, the values of the energy channels become $\omega_- = 0.026$ and $\omega_+ = 1.013$, indicating that $\omega_- \neq \omega_+$ (such that $\omega_+\gg\omega_-$). This difference in energy scales between the two channels enables the ancilla qubit to transfer additional temperature information, allowing for  more precise control over the operational range of temperature measurement at low temperatures. This is reflected in Fig.~\ref{fig5}(a), where we observe an extra peak in the QFI as a function of $T$. Since the low and high-temperature peaks in the QFI are associated with \( \omega_- \) and \( \omega_+ \), respectively, the corresponding maxima are determined with a precision of up to three decimal places. These values are given by $T_- = \omega_-/4$, and $T_+ = \omega_+/4.4$.

The physical mechanism behind the energy transitions induced by the bath, responsible for possible energy channels, can be further explained using Fig.~\ref{fig2}. The coupling parameters \( g \) and \( J \) play a crucial role in determining the behavior of transition energies. For two qubits, the transition energies are given by \( \omega_1 = \omega_S - \eta \) and \( \omega_2 = \omega_S + \eta \) (where $\omega_1$ and $\omega_2$ correspond to $\omega_-$ and $\omega_+$, respectively), as shown in Fig.~\ref{fig2}. Both frequencies depend on \( g \) and \( J \), with the key quantity being the energy splitting \( \Delta\omega = \omega_1 - \omega_2 \). This splitting makes the transition frequencies distinct and directly influences the second peak of the QFI.

Increasing either \( g \) or \( J \) modifies \( \Delta\omega \), causing the transition energies to converge. As a result, the QFI exhibits a single peak at a lower temperature, while the peak at higher temperatures diminishes. However, at very strong coupling strengths, the behavior may differ, necessitating careful analysis of transition energies. To ensure that transition energies remain sufficiently distinct, optimizing these coupling parameters is crucial. Therefore, it is essential to have multiple peaks in QFI; the transition energies must be in different order.

\subsubsection{Approximate QFI}
To derive the analytical expressions for the QFI corresponding to each peak, we can use the approximate values of the probe's population. Therefore, the QFI can be computed based on the probe population in Eq.~(\ref{popL}), and using the formula in Eq.~(\ref{qfi}), it is given by
\begin{equation}
    \mathcal{F}_Q^\text{low}\approx\frac{\omega_-^{2}\cos ^2(\theta ) e^{2\omega_-/T}}{T^4 \left(1+e^{\omega_- /T}\right)^2 \left(\sin ^2(\theta )+e^{\omega_-/T}\right)}.
\end{equation}
\begin{figure}[t!]
    \centering
    \includegraphics[scale=0.6]{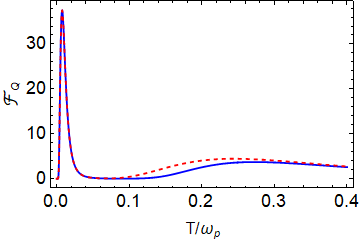}
    \caption{QFI $\mathcal{F}_Q$ as a function of $T$ for $\omega_p=1$, $\omega_a=0.04$, $g=0.02$, and $J=0.04$. The solid blue curves are obtained using exact QFI in Eq.~\eqref{exact}, and the red dashed curve is plotted using approximate QFI in Eq.~(\ref{app-qfi}). All the system parameters are scaled with the probe qubit frequency $\omega_p=1$.}
    \label{fig6}
\end{figure}
In a similar way, we apply the assumptions for the high temperature limit; the expression of QFI using Eq.~(\ref{popH}) for high temperature limit is given by
\begin{equation}
 \mathcal{F}_Q^\text{high}\approx\frac{\omega_+^2 \text{sech}^2\left(\frac{\omega_+}{2 T}\right)}{4 T^4}.\label{approx}
\end{equation}
The total QFI is the sum of these two and is given by
\begin{equation}\label{app-qfi}
    \mathcal{F}^\text{approx}_Q\approx \mathcal{F}_Q^\text{low}+\mathcal{F}_Q^\text{high}.
\end{equation}
Both the exact (Eq.~(\ref{exact})) and approximate (Eq.~(\ref{approx})) QFI values as a function of \(T\) are shown in Fig.~\ref{fig6}. In general, they agree well, especially in the low-temperature regime, where they are identical. A small difference appears at high temperatures.
 We give  a detailed discussion of the classical Fisher information and optimal measurements in Appendix~\ref{QFI}.
\section{Multiple ancilla qubits}\label{mq}
 In our study of the two-qubit scenario, we observed that the QFI peaks are closely related to the allowed transitions within the system. Since there are only two possible transitions for the two-qubit case, the system can exhibit at most two QFI peaks. In the following section, we extend our analysis to a larger chain of $N$ qubits, which enables the measurement of a broader range of temperatures.
 \subsection{Probe state}
The Hamiltonian for this system, which incorporates the interactions of interconnected multiple ancilla qubits with the probe, can be expressed as follows:
\begin{align}\label{mQ}
    \hat{H}&=\sum_{i=1}^{N}\frac{1}{2}\omega_{i}\hat{\sigma}^z_{i}+\sum_{i=1}^{N-1} J_i  \left(\hat{\sigma}^x_i\hat{\sigma}^x_{i+1}+\hat{\sigma}^y_i\hat{\sigma}^y_{i+1}\right)\nonumber \\ &+\sum_{i=1}^{N-1}g_i\left(\hat{\sigma}^x_i\hat{\sigma}^y_{i+1}-\hat{\sigma}^y_i\hat{\sigma}^x_{i+1}\right).
\end{align}
with the probe being the qubit numbered `$N$' so $\omega_N=\omega_p$ and $J_N$ and $g_N$ denote the coupling of the probe to the system via the last ancilla qubit.

To investigate the case of multiple ancilla qubits analytically, we find a reduced state for the probe qubit in the case of a chain of $N$ qubits, which is given by (see Appendix~\ref{Nqubits} for detailed calculations)
\begin{align} \label{SSden}
\hat{\rho}_p= \frac{Tr_A[e^{-\beta\sum_{ij}M_{ij}\hat{c}_i^\dagger\hat{c}_j}]}{\Pi_{l=1}^{N}(1+e^{-\beta E_l})},
\end{align}
where \(\hat{c}_i^\dagger (\hat{c}_i)\) are the fermionic creation (annihilation) operators corresponding to the \(i^{\text{th}}\) qubit, obtained through the Jordan-Wigner transformation, and $Tr_A$ signifies the trace over the $N-1$ ancilla fermions and $A=\{1,2,\ldots,N-1\}$. The partition function, $\mathcal{Z} = \prod_{l=1}^N (1 + e^{-\beta E_l})$, encapsulates the information about all the transition frequencies \(E_l\) present in the system. 
\par  As detailed in Appendix~\ref{Nqubits}, for a system with $N$ qubits, the maximum number of distinct transition frequencies is $N$. Since the partition function encodes information about these transition frequencies, the probe state can access this information, allowing for the detection of a wide range of temperatures through local measurements of the probe qubit.

The transition frequencies can be determined by calculating the eigenvalues of the $N \times N$ tridiagonal matrix $M$, as described below (see Appendix \ref{Nqubits}),
 \begin{align} \label{M_Matrix}
     M_{i,i}&=\omega_i, \nonumber\\
     M_{i,i+1}&=2(J_i+i g_i), \nonumber\\
     M_{i+1,i}&=2(J_i- i g_i),
 \end{align}
 with $M_{i,j} = 0$ for all other combinations of $i$ and $j$. The transition frequencies $E_l$ that appear in the partition function correspond to the eigenvalues of this matrix.  It becomes evident from this analysis that the maximum number of QFI peaks we can observe increases linearly with the system size, as the transition frequencies correspond to the eigenvalues of the \( N \times N \) $M$ matrix.
 
\subsection{Absence of coherences} Now, since we are focusing solely on the transition frequencies, it is essential to prove that coherences do not play a role in our system.
 It might be more prudent to evaluate the thermal state density matrix in the energy basis of the fermionic Hamiltonian, where it is just given as (see Appendix \ref{Nqubits}),
 \begin{equation}
    \hat{\rho}_\text{th}=\frac{e^{-\beta\left(\sum_lE_l\hat{a}_l^\dagger\hat{a}_l\right)}}{\mathcal{Z}}.
\end{equation}
After doing some algebra, we can prove that this matrix can be written in the original basis as,
\begin{equation} \label{simplied_form}
    \hat{\rho}_{th}=\frac{{\LARGE \Pi}_k \left(1+\sum_{i,j}(e^{-\beta E_k} -1) g\hat{\text{U}}_{kj}^\dagger\hat{\text{U}}_{ik}\hat{c}_i^\dagger \hat{c}_j\right)}{\mathcal{Z}},
\end{equation}
where $\hat{\text{U}}$ is the unitary matrix that diagonalizes the system Hamiltonian with details mentioned in Appendix~\ref{Nqubits}.
 Looking at the above form of the density matrix, we try to establish that there will be no coherences in the probe state.
  
\begin{align}
    \hat{\rho}_p = \frac{Tr_A \left [{{\LARGE \Pi}_k\left(1+\sum_{i,j}(e^{-\beta E_k} -1) \hat{\text{U}}_{kj}^\dagger\hat{\text{U}}_{ik}\hat{c}_i^\dagger \hat{c}_j\right)} \right ]}{\mathcal{Z}}.
\end{align}  
Suppose we want to evaluate the matrix element $\langle 0|\hat{\rho}_p|1 \rangle$, which measures the coherence in the probe state. We also define $\theta_k^{i,j}=(e^{-\beta E_k} -1) \hat{\text{U}}_{kj}^\dagger\hat{\text{U}}_{ik}$ for notational convenience,
\begin{align}\label{elm}
    \langle 0|\hat{\rho}_p|1 \rangle = \frac{1}{\mathcal{Z}} \langle 0|Tr_A \left [{ {\LARGE \Pi}_k \left(1+\sum_{i,j}\theta_k^{i,j}\hat{c}_i^\dagger \hat{c}_j\right)} \right ]|1 \rangle.
\end{align}  
The only way the term in question can be non-zero is if it involves both the fermionic operator $\hat{c}_N$ and its Hermitian conjugate in the probe's density matrix. The thermal state, as defined in Eq.~\eqref{simplied_form}, is a complex many-body operator. When fully expanded, this operator consists of a sum of terms involving different numbers of fermionic operators. These terms include:

\begin{itemize}
    \item 2-fermion terms, such as $\hat{c}_1^\dagger \hat{c}_2$,
    \item 4-fermion terms, such as $\hat{c}_1^\dagger \hat{c}_2 \hat{c}_3^\dagger \hat{c}_4$,
    \item 8-fermion terms, such as \[
\hat{c}_1^\dagger \hat{c}_2 \hat{c}_3^\dagger \hat{c}_4 \hat{c}_5^\dagger \hat{c}_6 \hat{c}_7^\dagger \hat{c}_8,
\] and so on, up to $2^N$-fermion terms.
\end{itemize}
For the probe density matrix to exhibit coherences, it must arise from a term containing only the raising or lowering operators of the probe fermion. For example, terms like $\hat{c}_i^\dagger \hat{c}_j \hat{c}_k^\dagger \hat{c}_N$ would contribute to coherence. 

However, since the partial trace over the ancillas corresponds to an expectation value taken with respect to the ancillary states, only those terms will survive the partial trace where both the raising and lowering operators of the same ancillary fermions are present.  This condition is not satisfied for operators such as $\hat{c}_i^\dagger \hat{c}_j \hat{c}_k^\dagger \hat{c}_N$, because these contain an odd number of ancillary fermionic operators. This results in having no coherences in the probe state.

Since coherence is absent in the probe state, the QFI depends solely on the population ($p(T)$). Given that the trace of the density matrix is \(1\), the QFI effectively depends on a single element of the probe density matrix, namely  $\langle 0|\hat{\rho}_p|0 \rangle$. We now try to write the form of this matrix element, such as 
\begin{align}\label{probe_function}
    \langle 0|\hat{\rho}_p|0 \rangle = \frac{1}{\mathcal{Z}} \langle 0|Tr_A \left [{ {\LARGE \Pi}_k \left(1+\sum_{i,j}\theta_k^{i,j}\hat{c}_i^\dagger \hat{c}_j\right)} \right ]|0 \rangle.
\end{align}  
Analyzing this term further analytically becomes challenging. Therefore, we shift to numerical investigations, carefully selecting parameters such that the transition frequencies $E_l$ are separated by distinct orders, leading to well-defined peaks in the quantum Fisher information.
 We have earlier seen in the two-qubit case that if all the transition frequencies are of distinct order, we can roughly estimate the Fisher peak points by solving the  transcendental equation, which is given below
\begin{equation}\label{mag}
   \tilde{T}_i\approx \frac{E_i}{2}\tanh\left(\frac{E_i}{2\tilde{T}_i}\right),
\end{equation}
 where, $E_i$ denotes the transition energies available in the system.  We assume that this equation can also serve as a guide in the case of multiple QFI peaks, provided that the eigenvalues of the \( `M' \) matrix, defined in \eqref{M_Matrix}, are of distinct orders. To verify this assumption, we choose parameters that ensure the eigenvalues differ in order and confirm its validity in the following subsections.

\subsection{$N=3$}
\begin{figure}[t!]
    \centering
    \includegraphics[scale=0.65]{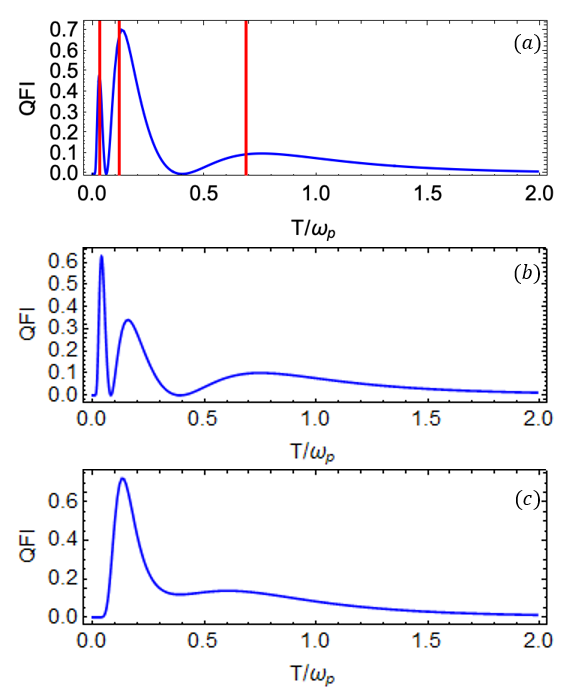}
    \caption{QFI of probe qubit as a function of temperature $T$ for the case of two ancilla qubits attached to the bath ($N_A=2$). Figures (a)-(c) correspond to plots for $g_1=0.04, 0.06, 0.1$, respectively. For each $g_1$ value we set $J_1=0.06, 0.08, 0.3$, respectively. The remaining parameters are fixed as $\omega_p=1$, $\omega_2=0.4$, $\omega_1=0.04$, $g_2=0.4$, and $J_2=0.4$. The solid red vertical lines indicate the temperature values calculated using Eq.~\eqref{mag} for the eigenvalues of the M matrix defined in Eq.~\eqref{M_Matrix}. All the system parameters are scaled with the probe qubit frequency $\omega_p=1$.}
    \label{fig8}
\end{figure}
It is worth noting that when the eigenvalues of the matrix \(M\) in Eq.~\eqref{M_Matrix} differ significantly in magnitude, the locations of the QFI peaks can be determined by numerically solving Eq.~\eqref{mag}. Our approximate numerical results indicate that each QFI peak reaches its maximum value at \(T = E_i / 2.4\).
\begin{figure*}[t!]
    \centering
    \subfloat[]{
    \includegraphics[scale=0.64]{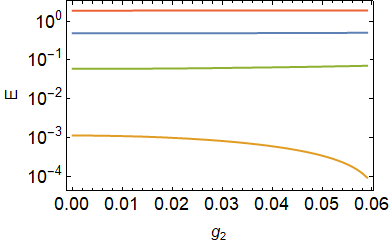}}
    \subfloat[]{
    \includegraphics[scale=0.64]{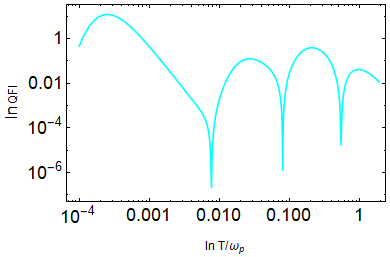}}\\
    \subfloat[]{
    \includegraphics[scale=0.64]{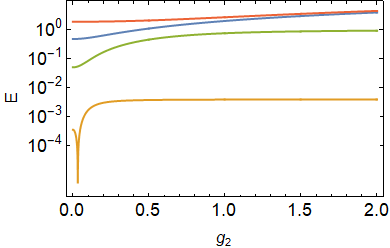}}
    \subfloat[]{
    \includegraphics[scale=0.64]{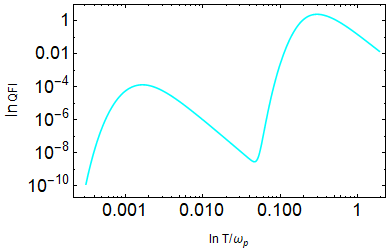}}
    \caption{ \textbf{Top row: } \textbf{(a)} Transition energies \( E \) as a function of coupling strength \( g_2 \) and \textbf{(b)} the corresponding QFI as a function of temperature $T$ for \( N=4 \), with the parameter set: $\omega_p=1$, $\omega_3=0.4$, $\omega_2=0.04$, $\omega_1=0.004$, $J_1=0.007$, $J_2=0.06$, $J_3=0.4$, $g_1=0.005$, and $g_3=0.4$. For plotting QFI, we considered weak coupling strength of $g_2=0.04$.
  \textbf{Bottom row:} Transition energies \( E \) as a function of coupling strength \( g_2 \) and the corresponding QFI as a function of temperature $T$ for \( N=4 \), with the parameter set: $\omega_p=1$, $\omega_3=0.4$, $\omega_2=0.04$, $\omega_1=0.004$, $J_1=0.006$, $J_2=0.04$, $J_3=0.4$, $g_1=0.004$, and $g_3=0.4$. For plotting QFI, we considered strong coupling strength of $g_2=2$.}
 \label{figT}
\end{figure*}
\par In what follows, we present our results for the case with more than one ancilla qubit. In particular, for $N_A=2$ (where $N_A$ represents the number of ancilla qubits, such as $N_A=N-1$), we plot the QFI for the probe qubit as a function of temperature \(T\) for different values of coupling strength when two ancilla qubits are attached to the thermal bath shown in Fig.~\ref{fig8}. The addition of a second ancilla qubit introduces another energy channel, allowing for enhanced information exchange between the bath and the probe. As a result, there are three energy channels in total: two previously established channels, which we explored in the previous section, and the new channel that enables us to investigate the low-temperature regime. This range enhancement is reflected as the emergence of a third peak in the QFI at low temperatures. The vertical lines in Fig.~\ref{fig8}(a) represent the values $\Tilde{T}_i$ calculated using Eq.~\eqref{mag} for the eigenvalues of the matrix in Eq.~(\ref{M_Matrix}) each corresponding to the respective QFI peak. These values are obtained by numerically solving the general form of Eq.~(\ref{mag}) using the same set of parameters.
In this case, we fixed the probe and ancilla frequencies as follows: $\omega_p=1$ (probe), $\omega_2=0.4$ (first ancilla), and $\omega_1=0.04$ (second ancilla). The coupling strengths of the $XX$ interaction $J$ are fixed, such as $J_2=0.4$ and $J_1=0.05$.
The coupling strength between the probe and first ancilla is set to $g_2=0.4$ and we vary the coupling strength $g_1$ between the probe and second ancilla. The heights of the two peaks at low temperatures can be adjusted by varying the coupling strengths $g_1$ and $g_2$, respectively. For example, increasing $g_1$ enhances the height of the third (leftmost) peak (as shown in Fig.~\ref{fig8}(c)), improving the precision of temperature estimation at lower temperatures. In Fig.~\ref{fig8}(c), we observe that the two peaks at higher temperature values have merged, making them indistinguishable. However, the precision of temperature estimation due to the middle peak decreases, and the height of this peak can be increased by tuning $g_2$ by fixing the value of $g_1$. We can conclude that there is a trade-off between parameters, such as $g_i$ and $J_i$, allowing us to fine-tune their values to enhance precision within the desired temperature range.
\begin{figure}[t!]
    \centering
    \includegraphics[scale=0.65]{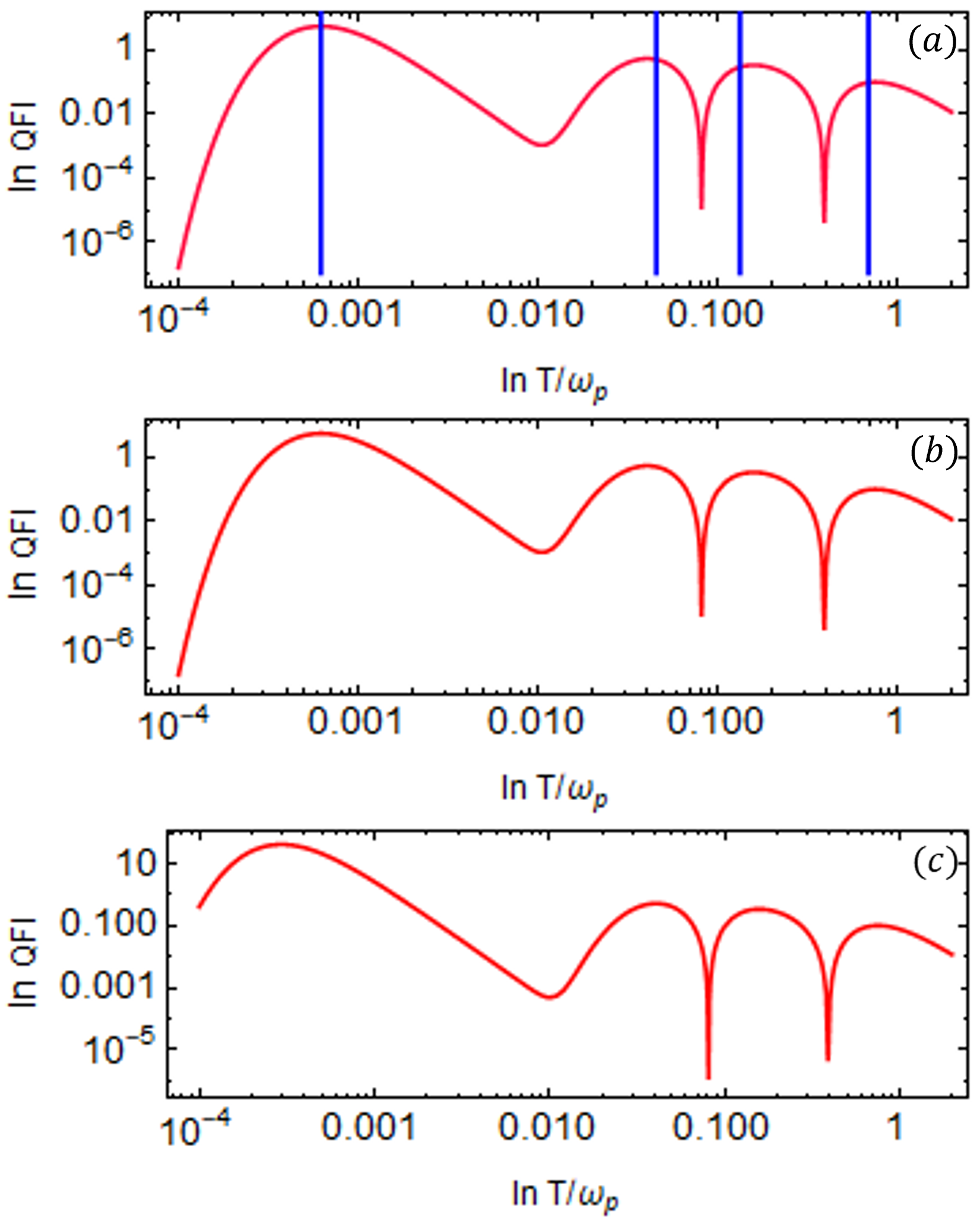}
    \caption{QFI of probe qubit as a function of temperature $T$ on a log-log scale is shown for the case of three ancilla qubits attached to the bath ($N_A=3$). Figures (a)-(c) correspond to the plots for $g_1 = 0.0055, 0.006, 0.0065$, respectively. For each $g_1$ value, we set $J_1 = 0.0075, 0.008, 0.0085$. respectively. The rest of the parameters are fixed to $\omega_p=1$, $\omega_3=0.4$, $\omega_2=0.04$, $\omega_1=0.004$, $g_3=0.4$, $g_2=0.06$, $J_3=0.4$, and $J_2=0.08$. The solid blue vertical lines indicate the temperature values calculated using Eq.~(\ref{mag}) for the eigenvalues of the matrix in Eq. ~(\ref{M_Matrix}) associated with each peak. All the system parameters are scaled with the probe qubit frequency $\omega_p=1$.}
    \label{fig9}
\end{figure}
\subsection{$N=4$}
We now extend the system to include three ($N_A=3$) ancilla qubits and present the results in Fig.~\ref{fig9} for different values of coupling strength $g_1$ between the probe and the third ancilla qubit. The QFI is plotted as a function of temperature $T$ on a log-log scale for different values of the coupling strength, $g_1$, and the other parameters are fixed. The inclusion of this third ancilla introduces an additional energy channel, allowing us to explore the low-temperature regime further. With four energy channels in total for the exchange of temperature information, the QFI exhibits four distinct peaks as a function of temperature $T$. Numerical analysis shows that the newly added ancilla should operate at a lower energy scale, with its transition frequency $\omega_1$ that should be set lower than both $\omega_2$ and $\omega_3$. Likewise, the coupling between the probe and the additional ancilla must remain weak, i.e., $g_1$ should be small.

We fixed the probe frequency at $\omega_p = 1$, and the ancilla qubit frequencies are set to $\omega_3 = 0.4$ (first ancilla), $\omega_2 = 0.04$ (second ancilla), and $\omega_1=0.004$ (third ancilla). The coupling strength between the probe and the first ancilla is set to $g_3 = 0.4$, and with the second ancilla, its value is $g_2=0.05$. We plot the QFI for different values of $g_1$, which represents the coupling strength between the probe and the third ancilla qubit. The corresponding Heisenberg coupling strengths $J$ are shown in Fig.~\ref{fig9}. To enhance precision in the low-temperature regime, $g_1$ must be properly tuned. We consider three different values: $g_1 = 0.0055$, $g_1 = 0.006$, and $g_1 = 0.0065$, as illustrated in Figs.~\ref{fig9}(a), (b), and (c), respectively. As $g_1$ increases, the height of the QFI peak in the low-temperature regime also increases, leading to improved temperature estimation precision.\\
Similarly, to enhance precision due to other QFI peaks, we can tune $g_2$ and $g_3$. Therefore, optimizing the parameters associated with specific QFI peaks is crucial for improving the accuracy of temperature measurements. Additionally, for distinct transition energies, the ancilla qubit transition frequencies should follow the hierarchy $\omega_1 < \omega_2 < \omega_3$, and the coupling strengths should satisfy $g_1 < g_2 < g_3$.
\begin{figure}[b!]
    \centering
    \includegraphics[scale=0.65]{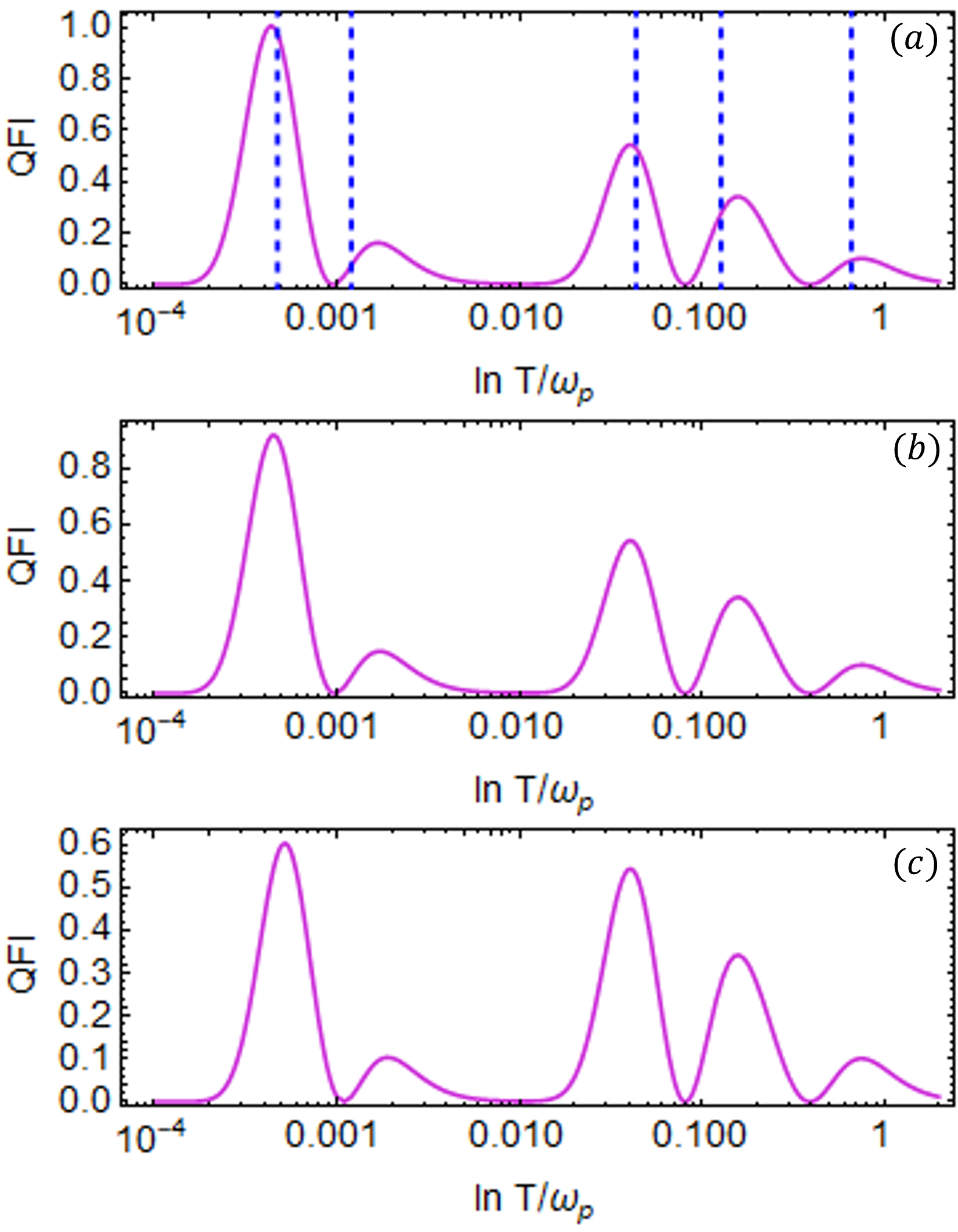}
    \caption{QFI of probe qubit as a function of temperature $T$ in the case of four ancilla qubits attached to the bath ($N_A=4$). Figures (a)-(c) represent the plots for $g_1 = 0.0005$, $g_1=0.00055$, and $g_1= 0.0007$, respectively. The rest of the parameters are fixed to $\omega_p=1$, $\omega_4=0.4$, $\omega_3=0.04$, $\omega_2=0.004$, $\omega_1=0.0004$, $g_4=0.2$ and $g_3=0.06$ , and $g_2=0.005$. The values of $J$ are fixed as follows: $J_4=0.4$, $J_3=0.08$, $J_2=0.008$, and $J_1=0.00095$. The dashed blue vertical lines indicate the temperature values associated with each peak. All the system parameters are scaled with the probe qubit frequency $\omega_p=1$.}
    \label{fig10}
\end{figure}

The reduced state of the probe given in Eq.~\eqref{SSden} is a many-body state; therefore, it is difficult to draw direct conclusions that which peaks of QFI correspond to the available transitions and establish the role of transition energies in resulting the QFI peaks. We can infer the parameter ranges where the QFI peaks align with the transition energies. To this end, we now discuss the influence of transition energies on the number of peaks in the QFI as a function of temperature. For multiple ancilla qubits, we numerically solve the general eigenvalue equation in Eq.~\eqref{M_Matrix} to determine the transition energies for any number of qubits. Figure~\ref{figT} illustrates the available transition energies $E$ as a function of $g_2$ obtained by solving Eq.~\eqref{M_Matrix} for $N=4$ and the corresponding QFI under different parameter choices.\\
When the onset frequencies of the qubits are distinct and follow a hierarchy, as discussed earlier, the transition energies as a function of $g_2$ remain well-separated and we have four frequencies that exhibit a clear ordering, as shown in Fig.~\ref{figT}(a). Consequently, the QFI displays distinct  four peaks corresponding to these $E$ values, as depicted in Fig.~\ref{figT}(b). This shows how the transition energies are aligned with the QFI peaks.\\
Conversely, if the ancilla qubits still have the same values but transition energies are plotted as a function of strong coupling values, such as $g_2$, we can still see different transition energies as shown in Fig.~\ref{figT}(c).\\
Interestingly, in this scenario, the QFI exhibits only two peaks as a function of temperature $T$. This occurs because increasing the coupling strength causes energy levels to converge, and there are only two frequencies for which QFI can be well separated, thereby reducing the number of peaks.\\
In general, interactions play a nuanced role, depending on the specific parameters considered. From these observations, we conclude that distinct ancilla qubit frequencies are not a strict requirement for multiple QFI peaks; rather, the existence of different transition energies primarily governs the number of peaks. A similar trend holds when fixing $g_2$ and plotting $E$ as a function of other coupling values.

\subsection{$N=5$}
Finally, we investigate the system by increasing the total number of ancilla qubits to  $N_A=4$. This introduces yet another energy channel, allowing us to probe even deeper into the low-temperature regime. As expected, with the addition of this new energy channel, the QFI now displays five distinct peaks as a function of temperature $T$, corresponding to the additional level of control over the system dynamics in the ultralow temperature regime. The results are plotted in Fig.~\ref{fig10} as a function of $T$ for different values of coupling strength $g_1$ between the probe and fourth ancilla. The introduction of a fourth ancilla adds a new energy channel that allows the probe to estimate the ultralow temperature regime, the precision of which can be enhanced by tuning $g_i$ values. In Fig.~\ref{fig10}(a)-(c), we fixed $g_2$, $g_3$, and $g_4$ while varying $g_1$ from $g_1=0.0005$ to $g_1=0.0007$. This adjustment increased the height of the third QFI peak (from the right) but reduced the precision at the low-temperature peak. Similarly, by fine-tuning other $g$ values over a broader range of measurable temperatures. This suggests that to improve estimation precision in the low-temperature regime, it should be set to $g_1\le0.0005$ while the other $g$ values can be optimized to enhance the precision of the remaining peaks. Thus, each $g$ value can be adjusted to modulate the height of its corresponding QFI peak. 

From these results, we find that adding more ancilla qubits introduces additional energy channels that facilitate the exchange of temperature information, thereby extending the accessible temperature range to lower regimes. The results suggest that the probe qubit should always be at a higher energy scale, and properly setting this energy scale difference is crucial for probing the low-$T$ regime. As seen in the two-qubit case, where there are only two energy channels, $\omega_-$ and $\omega_+$, the QFI shows two peaks. For a higher number of ancilla qubits, we can determine the jump frequencies and predict the location of the peaks in the QFI through $E_l$. For instance, in the case of $N_A=4$, the order should be $\omega_1 < \omega_2 < \omega_3 < \omega_4 < \omega_p$, and similarly, the coupling strengths should follow $g_1 < g_2 < g_3 < g_4$. By following a similar analysis, one can identify an arbitrary number of peaks in the QFI. However, as all these peaks occur at low temperatures, we may observe smooth behavior in the QFI for a large number of ancilla qubits. Our results indicate a significant method for measuring ultralow temperatures using chains of qubits. Moreover, while increasing the number of ancilla qubits extends the range of measurable temperatures, it also leads to a reduction in the precision of temperature estimation.
\section{Conclusion}\label{conc}
In this paper, we presented a scheme for estimating low temperatures using a qubit chain coupled via Heisenberg $XX$ and DM interactions. We exploit the allowed transitions between energy levels to measure low temperatures in quantum thermometry. We first consider a two-qubit configuration where the ancilla qubit is connected to both the thermal bath and the probe and serves as a mediator. The dynamics of this open quantum system are described by a Gibbs thermal state under the assumption of weak system-bath coupling. 
We first investigated the sensitivity of the populations of the probe qubit with respect to temperature under two distinct cases. For the resonant qubits, we found that the first derivative of the excited state of the population increases as a function of temperature \(T\), exhibiting a single peak. Conversely, in the off-resonant case, we observed an additional peak with a smaller amplitude and sharper variations at lower temperatures. This behavior arises because there are two different energy transitions, \(\omega_+\) and \(\omega_-\) (where \(\omega_+ \neq \omega_-\)), which facilitate the exchange of temperature information between the bath and the probe. In the resonant case \(\omega_+ \approx \omega_-\), this effectively reduces the system to a single transition. These findings emphasize that for the efficient low-temperature measurements, the two qubits should operate at different energy scales, with the probe qubit having a higher energy \((\hbar\omega_p > \hbar\omega_a)\) than the ancilla qubit.

We derived analytical expressions for the population of the excited state of the probe qubit and the corresponding temperature in both low- and high-temperature regimes, focusing on points where the first derivative of the population reaches its maximum. Using these approximate expressions of the population, we calculated the QFI separately for each regime and observed that the QFI as a function $T$ exhibits two peaks, corresponding to the two allowed transitions in the system.

Moreover, since the system does not generate coherences, all temperature information can be extracted solely from the populations of the probe, which adds another layer of robustness to the results. As a consequence, quantum measurements are not essential, underscoring the significance of this study. To validate this, we calculated the CFI for the probe and found it to be identical to the QFI. Subsequently, we explored optimal measurements that are capable of saturating the quantum Cramér-Rao bound. We observed that the Fisher information based on \(\hat{\sigma}^z\) provides the same temperature information as the QFI and CFI based on the population measurement. In contrast, the Fisher information based on \(\hat{\sigma}^x\) measurements yields zero information, which indicates that these measurements are not suitable for temperature estimation in the current scheme.

Finally, we investigated a chain of \(N\) interacting qubits. We derived an expression for the probe qubit that accounts for all energy channels (transitions) through \(E_l\). With \(N-1\) ancilla qubits, we can now have \(N + 1\) energy channels that facilitate the exchange of temperature information between bath and probe and imprint them on the state of the probe. These additional energy channels enhance the range of measurable temperatures, with each new peak in the QFI appearing progressively at lower temperatures. Furthermore, the precision associated with each QFI peak can be adjusted by fine-tuning the relevant system parameters. Therefore, with such a configuration, we can achieve an arbitrary number of peaks in the QFI associated with each transition, enabling the measurement of ultralow temperatures in quantum systems using  qubits as probes.

Our proposed thermometry scheme, based on physical interactions between qubits, provides a promising foundation for practical thermometric applications, offering enhanced sensitivity in ultra-low temperature regimes. By incorporating such natural interactions, the proposed scheme not only broadens the range of measurable temperature but also ensures experimental feasibility, bridging the gap between theoretical advances and their practical deployment in quantum thermometry.
\section*{Acknowledgment}
This work is supported by the Scientific and Technological Research
Council of Türkiye (TÜBİTAK) under Project Grant Number 123F150.
\appendix
\begin{widetext}
\section{Two-qubit density matrix}\label{trans}
Here, we would like to discuss the thermalization of two qubits with the thermal sample.  To this end, we consider weak system-bath coupling of the form given by Eq.~(\ref{sb}), allowing the entire two-qubit state to reach a Gibbs thermal steady
state~\cite{Breuer,lidar2020lecturenotestheoryopen}. 
The resulting steady state of the two qubits in diagonal basis is given by
\begin{equation}\label{rho_dia}
\Tilde{\rho}_{ss}=\frac{1}{\mathcal{Z}}\left(
\begin{array}{cccc}
 e^{-\beta\omega_S} & 0 & 0 & 0 \\
 0 & e^{-\beta\eta} & 0 & 0 \\
 0 & 0 & e^{\beta\eta} & 0 \\
 0 & 0 & 0 & e^{\beta\omega_S} \\
\end{array}
\right),
\end{equation}
where $\mathcal{Z}=2\big(\cosh{(\beta\omega_S)}+\cosh{(\beta\eta})\big)$ is the partition function expressed in the diagonal basis and $\beta=1/T (k_B=1)$ is the inverse temperature.
The density matrix for the two qubits in the computational basis can be obtained by applying the unitary transformation, which reads as
\begin{equation}\label{ss}
\hat{\rho}_{ss}=\hat{\text{U}}^\dagger\Tilde{\rho}_{ss}\hat{\text{U}}=\frac{e^{-\beta\hat{H}_S}}{\mathcal{Z}},
\end{equation}
where the explicit form of the unitary transformation $\hat{\text{U}}$ that is used to transform the density matrix $\Tilde{\rho}_{ss}$ from the diagonal basis to the computational basis is given by
\begin{equation}\label{ss}
\hat{\text{U}}=\left(
\begin{array}{cccc}
 1 & 0 & 0 & 0 \\
 0 & \frac{2(J+ig)}{\Delta} & \frac{\omega_D-\eta}{\Delta} & 0 \\
 0 & \frac{\eta-\omega_D}{\Delta} & \frac{2(J-ig)}{\Delta} & 0 \\
 0 & 0 & 0 & 1 \\
\end{array}
\right).
\end{equation}
In the two-qubit computational basis $\{|++\rangle,|+-\rangle,|-+\rangle,|--\rangle\}$, the density matrix $\hat{\rho}_{ss}$ is given by a two-qubit $X$ state \cite{PhysRevE.104.054137},
\begin{equation}\label{ss}
\hat{\rho}_{ss}=\left(
\begin{array}{cccc}
 \text{d}_1 & 0 & 0 & 0 \\
 0 & \text{d}_2 & c & 0 \\
 0 & c^* & \text{d}_3 & 0 \\
 0 & 0 & 0 & \text{d}_4 \\
\end{array}
\right),
\end{equation}
where the diagonal elements (populations) of the density matrix are given by
\begin{equation}\label{pop}
    \begin{aligned}
     d_1=&\frac{1}{2 e^{\omega_S/T} \cosh \left(\frac{\eta }{T}\right)+e^{2 \omega_S/T}+1},\quad
     d_2=\frac{\cos ^2(\theta )+\sin ^2(\theta ) e^{2 \eta/T}}{2 e^{\eta /T} \cosh \left(\frac{\omega_S}{T}\right)+e^{2 \eta /T}+1},\\
     d_3=&\frac{\sin ^2(\theta )+\cos ^2(\theta ) e^{2 \eta/T}}{2 e^{\eta /T} \cosh \left(\frac{\omega_S}{T}\right)+e^{2 \eta/T}+1},\quad
     d_4=\frac{e^{\omega_S/T}}{2 \chi}.
    \end{aligned}
\end{equation}
The off-diagonal term (coherence) is given by
\begin{equation}\label{coh}
    c=\frac{i\sin (2\theta ) \sinh \left(\frac{\eta }{T}\right)}{2\chi}.
\end{equation}
The mixing angle $\theta$ is defined as 
\begin{equation}\label{ma}
    \theta=\sin ^{-1}\left(\frac{\omega_D-\eta }{\sqrt{4 g^2+4 J^2+(\omega_D-\Omega )^2}}\right).
\end{equation}
In Eqs.~(\ref{pop}) and (\ref{coh}), we introduced a parameter that reads as 
\begin{equation}
    \begin{aligned}
        \chi&=\cosh \left(\frac{\eta }{T}\right)+\cosh \left(\frac{\omega_S}{T}\right).
    \end{aligned}
\end{equation}
\section{Classical Fisher information and optimal measurements}\label{QFI}
In our analysis, we initially examined the temperature-dependent populations, which play a pivotal role in temperature estimation. The temperature sensitivity was further quantified using the QFI as a key figure of merit, providing a rigorous measure of the precision achievable in temperature measurements. Given that the system lacks coherences, we can now use the classical Fisher information (CFI) to quantify the amount of information that the temperature-dependent populations provide regarding the temperature estimation. The CFI serves as a valuable measure of sensitivity in parameter estimation, allowing us to assess how well the populations can be utilized to infer the unknown temperature of the system.
The general form of the CFI for a probability distribution $p(x;\theta)$ with respect to a parameter $\theta$ can be expressed as~\cite{PhysRevLett.72.3439,paris2009}
\begin{equation}\label{cfi}
    \mathcal{I}(\theta)=\int\frac{1}{p(x;\theta)}\left(\frac{\partial}{\partial\theta}\ln{p(x;\theta)}\right)^2dx,
\end{equation}
where $p(x;\theta)$ denotes the probability distribution of $\theta$ as a function $\theta$.
We calculated the CFI by using the temperature-dependent populations of the probe's density matrix (using Eq.~(\ref{cfi})). Remarkably, we find that the CFI is exactly equal to the QFI, which we computed earlier using Eq.~(\ref{qfi}). This equivalence indicates that, in this case, classical and quantum measurements converge, and the CFI captures the same amount of information as the QFI for temperature estimation, i.e., $\mathcal{I}(T) = \mathcal{F}_Q$. This result highlights that the classical measurement scheme based on population measurement alone can be as efficient as a full quantum approach, reinforcing the practical relevance of using population measurements in quantum thermometry.

It is essential to determine the best optimal measurement that can saturate the quantum Cramer-Rao bound. Therefore, we will present a measurement protocol that is implemented in practical experiments. 
For a single qubit thermometer, we chose projective measurements, and the corresponding Fisher information is given by~\cite{PhysRevLett.125.080402}
\begin{equation}
    \mathcal{F}(\langle\hat{X}\rangle) = \frac{1}{\text{Var}(\hat{X})} \left( \frac{\partial \langle \hat{X} \rangle}{\partial T} \right)^2, \label{qfi-sz}
\end{equation}
where $\langle\hat{X}\rangle$ and $\langle\Delta\hat{X}\rangle$ denote the mean and variance of of the observable $\hat{X}$, respectively. 
\begin{figure}[t]
    \centering
    \includegraphics[scale=0.6]{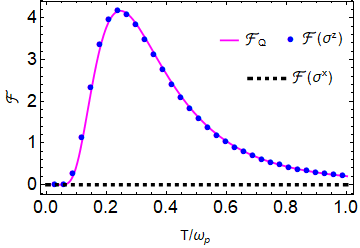}
    \caption{Fisher information calculated using Eq.~(\ref{qfi}) and the optimal Fisher information based on $\sigma^z$ measurement are plotted as a function of temperature $T$. The parameters are set to $\omega_p=1$, $\omega_a=1$, $g=0.02$, and $J=0.04$. All the system parameters are scaled with the probe qubit frequency $\omega_p=1$.}
    \label{fig7}
\end{figure}

To compute the Fisher information based on measurement $ \langle\hat{\sigma}^z\rangle$, we employ the probe state and derive the following quantities:
\begin{equation}
\begin{aligned}
    \langle\hat{\sigma}^z\rangle=&\frac{\cos (2 \theta ) \sinh \left(\frac{\eta }{T}\right)-\sinh \left(\frac{\text{$\omega $s}}{T}\right)}{\cosh \left(\frac{\eta }{T}\right)+\cosh \left(\frac{\text{$\omega $s}}{T}\right)},\\
    \langle\Delta\hat{\sigma}^z\rangle&=1-\frac{\left(\sinh \left(\frac{\text{$\omega $s}}{T}\right)-\cos (2 \theta ) \sinh \left(\frac{\eta }{T}\right)\right)^2}{\left(\cosh \left(\frac{\eta }{T}\right)+\cosh \left(\frac{\text{$\omega $s}}{T}\right)\right)^2}.
    \end{aligned}
\end{equation}
Using these expressions along with Eq.~(\ref{qfi-sz}), we get the following form of $\langle\hat{\sigma}^z\rangle$ based Fisher information,
\begin{equation}\label{qfiz}
    \mathcal{F}(\langle\hat{\sigma}^z\rangle)=\frac{\left(\mathcal{B}+\zeta +\eta  \cos (2 \theta ) \sinh ^2\left(\frac{\eta }{T}\right)-\eta  \cos (2 \theta ) \cosh ^2\left(\frac{\eta }{T}\right)+\text{$\omega $s}\right)^2}{T^4(1-\alpha ) \chi^4},
\end{equation}
where the parameter $\alpha$ is given by 
\begin{equation}
\begin{aligned}
    \alpha:=&\frac{\left(\sinh \left(\frac{\text{$\omega $s}}{T}\right)-\cos (2 \theta ) \sinh \left(\frac{\eta }{T}\right)\right)^2}{\chi^2}.
    \end{aligned}
\end{equation}
However, the QFI obtained from $\hat{\sigma}^x$ measurements is zero.
The QFI in Eqs.~(\ref{qfi}) and ~(\ref{qfiz}
) are plotted in Fig.~\ref{fig7} as a function of temperature $T$. It is observed that both methods ($\mathcal{F}_Q$ \text{and} $\mathcal{F}(\sigma^z)$) yield identical Fisher information, a result that holds across all parameter regimes. This demonstrates that the QFI is consistent with the optimal Fisher information. Importantly, the QFI derived from the $\sigma^x$ measurement is zero, highlighting its inefficacy in this context. These results suggest that precise temperature estimation at the quantum limit can indeed be achieved, but through carefully tailored measurement strategies that target the most informative observables, such as the $\sigma^z$ measurement for probes interacting with thermal samples.
\section{Chain of  $N$ qubits}\label{Nqubits}
The Hamiltonian for a chain of $N$ qubits in Eq.~(\ref{mQ}) can be transformed to a fermionic chain using the following Jordan-Wigner transformation
\begin{equation}
\begin{aligned}
    \hat{\sigma}^z_i=2\hat{\sigma}^+_i\hat{\sigma}^-_i-1, \quad \hat{\sigma}^x_i= \hat{\sigma}^+_i+\hat{\sigma}^-_i,\quad \hat{\sigma}^y_i= -i (\hat{\sigma}^+_i-\hat{\sigma}^-_i),\quad \hat{\sigma}^\pm_i= \frac{\hat{\sigma}^x_i \pm i\hat{\sigma}^y_i}{2}.
\end{aligned}
\end{equation}
Therefore, the the above Hamiltonian becomes
\begin{equation}
    \begin{aligned}
       \hat{\mathcal{H}}&=\sum_{i=1}^{N}\frac{\omega_i}{2}(2\hat{\sigma}^+_i\hat{\sigma}^-_i-1)-\sum_{i=1}^{N-1} i g_i\left[(\hat{\sigma}^+_i+\hat{\sigma}^-_i)(\hat{\sigma}^+_{i+1}-\hat{\sigma}^-_{i+1})-(\hat{\sigma}^+_{i+1}+\hat{\sigma}^-_{i+1})(\hat{\sigma}^+_i-\hat{\sigma}^-_i)
       \right] \nonumber\\& + \sum_{i=1}^{N-1} J_i\left[(\hat{\sigma}^+_i+\hat{\sigma}^-_i)(\hat{\sigma}^+_{i+1}+\hat{\sigma}^-_{i+1})+(\hat{\sigma}^+_i-\hat{\sigma}^-_i)(\hat{\sigma}^+_{i+1}-\hat{\sigma}^-_{i+1})
       \right] 
    \end{aligned}
\end{equation}
After doing some algebra for simplification, the above Hamiltonian takes the following form, given below
\begin{equation}
    \hat{\mathcal{H}}=\sum_{i=1}^{N}\omega_i\hat{\sigma}^+_i\hat{\sigma}^-_i +2\sum_{i=1}^{N-1} (J_i+ig_i) \hat{\sigma}^+_{i}\hat{\sigma}_{i+1}^-+(J_i-ig_i) \hat{\sigma}^+_{i+1}\hat{\sigma}_i^- 
\end{equation}
According to the Jordan-Wigner transformation, we can transform the spin operators into corresponding fermionic operators using 
\begin{equation}
\begin{aligned}
\hat{\sigma}_i^+ &= \hat{c}_i^\dagger e^{i\pi \sum_{l < n} \hat{c}_i^\dagger\hat{c}_i}, \\
\hat{\sigma}_i^- &= \hat{c}_i e^{-i\pi \sum_{l < n} \hat{c}_i^\dagger\hat{c}}.
\end{aligned}
\end{equation}
After doing some straightforward algebra, we get
\begin{equation}
\hat{\mathcal{H}} = {C}^\dagger \hat{M} {C},
\end{equation}
where we have ignored the scale changing by the constant term $-\sum_{i=1}^{N+1}\frac{\omega_i}{2}$ in the above Hamiltonian, as it will eventually cancel out when we consider the thermal state, and the vectors ${C}$ and ${C}^\dagger$ are given by
\begin{equation}
\begin{aligned}
{C} = \begin{pmatrix}
\hat{c}_1 \\
\hat{c}_2 \\
\hat{c}_3 \\
\vdots \\
\hat{c}_{N}
\end{pmatrix}, \quad {C}^\dagger = \begin{pmatrix}
\hat{c}^\dagger_1 & \hat{c}^\dagger_2 & \hat{c}^\dagger_3 & \dots & \hat{c}^\dagger_{N}
\end{pmatrix},
\end{aligned}
\end{equation}
and $\hat{M}$ is a $(N+1)\times (N+1)$ matrix that connects th fermionic operators $\hat{c}$ and $\hat{c}^\dagger$.
Hence
\begin{equation}
\hat{\mathcal{H}}={C}^\dagger \hat{M} {C},
\end{equation}
Now, suppose that we can diagonalize $\hat{M}$ by using the following transformation
\begin{equation}
    \Tilde{M}=\hat{\text{U}}^\dagger\hat{M}\hat{\text{U}}, \quad \text{which implies that} \quad \hat{M}=\hat{\text{U}}\Tilde{M}\hat{\text{U}}^\dagger,
\end{equation}
where $\hat{\text{U}}$ is a diagonalizing matrix and $\Tilde{M}$ is a diagonal matrix. Suppose that $(\Tilde{M})_{ij}=E_i\delta_{ij}$ and $\hat{M}_{in}=\hat{\text{U}}_{ij}\Tilde{M}_{jk}\hat{\text{U}}_{kn}^\dagger=\hat{\text{U}}_{ij}E_j\delta_{jk}\hat{\text{U}}_{kn}^\dagger$, which implies that $\hat{M}_{in}=\sum_j\hat{\text{U}}_{ij}E_j\hat{\text{U}}_{jn}^\dagger$. Using this in the Hamiltonian, we can write as 
\begin{equation}
\begin{aligned}
    \hat{\mathcal{H}} =& {C}^\dagger\hat{M}{C}\\
    &=\sum_{jk}\hat{c}_j^\dagger\hat{M}_{jk}\hat{c}_k\\
    &= \sum_{jkl}\hat{c}_j^\dagger\hat{\text{U}}_{jl}E_l\hat{\text{U}}_{lk}^\dagger\hat{c}_k.
\end{aligned}
\end{equation}
Defining the operators,
\begin{align}\label{operator_relation}
		\hat{a}_k&=\sum_j U^\dagger_{k,j} \hat{c}_j ,&&
		\hat{a}^{\dagger}_k=\sum_j U_{j,k} \hat{c}^{\dagger}_j,
     \end{align}   
we can write the Hamiltonian as
\begin{equation}
    \hat{\mathcal{H}}=\sum_lE_l\hat{a}_l^\dagger\hat{a}_l,
\end{equation}
where $E_l$ is the transition frequency available for the $l^{th}$ level. As we see that the steady-state solution of the two qubits in global basis is a thermal state, therefore, for the chain of $N$ qubits, we can write the thermal state as 
\begin{equation}
    \Tilde{\rho}_{th}=\frac{e^{-\beta\left(\sum_lE_l\hat{a}_l^\dagger\hat{a}_l\right)}}{\text{Tr}[e^{-\beta\left(\sum_lE_l\hat{a}_l^\dagger\hat{a}_l\right)}]}.
\end{equation}
where the partition function $\mathcal{Z}=\text{Tr}[e^{-\beta\sum_lE_l\hat{a}_l^\dagger\hat{a}_l}]$ can easily be found and is given by
\begin{equation}
    \mathcal{Z}=\Pi_{l=1}^{N}\left(1+e^{-\beta E_l}\right).
\end{equation}
This represents the thermal state of a chain of ancilla qubits, including the probe qubit, in the diagonal basis. To obtain the density matrix of the probe qubit, we first need to transform this thermal state back to the computational basis. After the transformation, we trace out all the ancilla qubits. Consequently, the density matrix in the computational basis is expressed as
\begin{equation}
    \hat{\rho}_{th}=\frac{e^{-\beta\sum_{ij}M_{ij}\hat{c}_i^\dagger\hat{c}_j}}{\mathcal{Z}}.
\end{equation}
The partition function in this case is basis-independent, and from its structure, it is evident that all the system's frequencies can be identified through the energy levels $E_l$. This indicates that the partition function encapsulates the complete spectrum of the system's energies, enabling the detection of all relevant frequencies. 
Finally, the state of the probe state in the computational basis in terms of the fermionic operators is given as,
\begin{align}
\hat{\rho}_p= \frac{1}{\mathcal{Z}}Tr_A[e^{-\beta\sum_{ij}M_{ij}\hat{c}_i^\dagger\hat{c}_j}],
\end{align}
where, $Tr_A$ signifies the trace over the $N-1$ ancilla fermions and $A=\{1,2,\ldots,N-1\}$.
\end{widetext}
\appendix
\bibliography{DM}
\end{document}